\documentclass[3p,review,number,sort&compress]{elsarticle}
\usepackage{lineno}
\usepackage{amsmath, amsthm, amssymb, graphicx, verbatim,subfigure,epstopdf}

\begin{document}

\title{Modeling phase-separated transition-edge sensors in SuperCDMS detectors}
\author[mit]{A.J. Anderson\corref{cor}}
\ead{adama@mit.edu}
\author[mit]{S.W. Leman}
\author[ucb]{T. Doughty}
\author[mit]{E. Figueroa-Feliciano}
\author[mit]{K.A. McCarthy}
\author[stanford]{M. Pyle}
\author[scu]{B.A. Young}

\address[mit]{Department of Physics, Massachusetts Institute of Technology, Cambridge, MA 02139}
\address[ucb]{Department of Physics, University of California, Berkeley, CA 94720}
\address[stanford]{Department of Physics, Stanford University, Stanford, CA 94305}
\address[scu]{Department of Physics, Santa Clara University, Santa Clara, CA 95053}

\cortext[cor]{Corresponding author}
\date{\today}

\begin{abstract}
The SuperCDMS experiment implements transition-edge sensors to measure athermal phonons produced by nuclear recoils in germanium crystals.  We discuss a numerical simulation of these TES devices and a procedure for tuning the free model parameters to data, which reproduces superconducting-to-normal phase separation within a TES.  This tuning provides insight into the behavior of the TESs, allows us to study the phase-separation length to optimize our detector design, and is integrated into a more complete simulation of the phonon and charge physics of SuperCDMS detectors.
\end{abstract}

\maketitle

\section{Introduction}
Transition-edge sensors (TESs) function as extremely sensitive calorimeters and are used in applications such as dark matter detection \cite{CDMSScience:2010,CRESST:2009}, x-ray astrophysics \cite{microx:2009,IXO:2010}, and measurements of the cosmic microwave background \cite{SPT:2004, SPIDER:2008}.  A TES consists of a superconducting material typically held under a constant voltage bias that maintains it in the superconducting transition.  Negative electrothermal feedback stabilizes the sensor against small perturbations in current or temperature \cite{IrwinAPL:1995}.  In this arrangement, a TES is a stable device that experiences a measurable current change in response to an energy input.  Applications of TESs exploit this behavior to achieve high energy sensitivity to small signals \cite{Irwin:2005}.

One technical complication that can arise when operating TESs is the occurrance of superconducting-to-normal phase separation within the device \cite{Leman:2006}.  Instead of being homogeneous, different parts of a TES may have different temperatures or thermodynamic phases.  This typically occurs when there is weak thermal coupling between different regions of the TES.  For example, the ends of a TES whose length is much greater than its width may be in different phases.

\begin{figure}
\begin{center}
\includegraphics[scale=0.5]{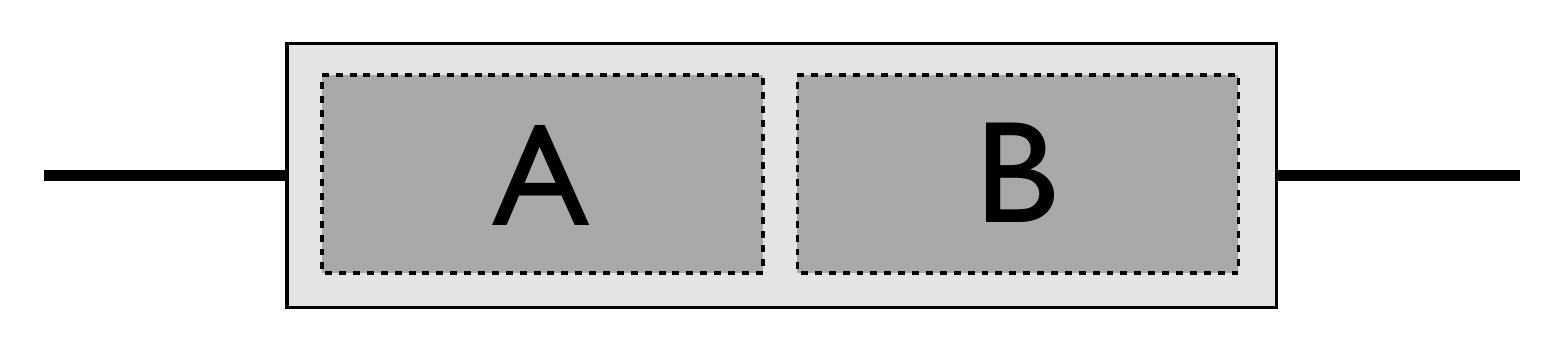}
\caption{If length of a TES is much longer than its width, then thermal diffusion between the two ends of the device can be very small.  In this situation, the TES behaves similar to two thermally disconnected TESs in series, and superconducting-normal phase-separation results.\label{fig:TwoPartTES}}.
\end{center}
\end{figure}

To heuristically understand why phase separation occurs, consider a long, voltage-biased TES, as shown in Figure \ref{fig:TwoPartTES}.  We can consider it as consisting of two thermally decoupled regions, denoted A and B.  If a thermal fluctuation causes the temperature of A to rise, then the resistance will also increase and, because the TES is voltage biased, the current will decrease.  This decreases the Joule heating and subsequently the temperature in B.  Region A thus tends toward the normal phase and region B tends toward the superconducting phase.  Thermal diffusion along the length of the TES counteracts this effect, but diffusion may be insufficient to prevent phase separation in a long TES.

Phase separation has several negative effects in applications.  The current response to a voltage or temperature impulse is different for phase-separated and phase-uniform cases.  Phase-separated TESs also have higher noise and reduced sensitivity \cite{AndersonLTD:2011}.  Internal thermal fluctuations become more significant because of the larger temperature gradient across the device.  These internal thermal fluctuations produce additional noise in TESs \cite{Galeazzi:2011}

\begin{figure}
\begin{center}
\includegraphics[scale=0.5]{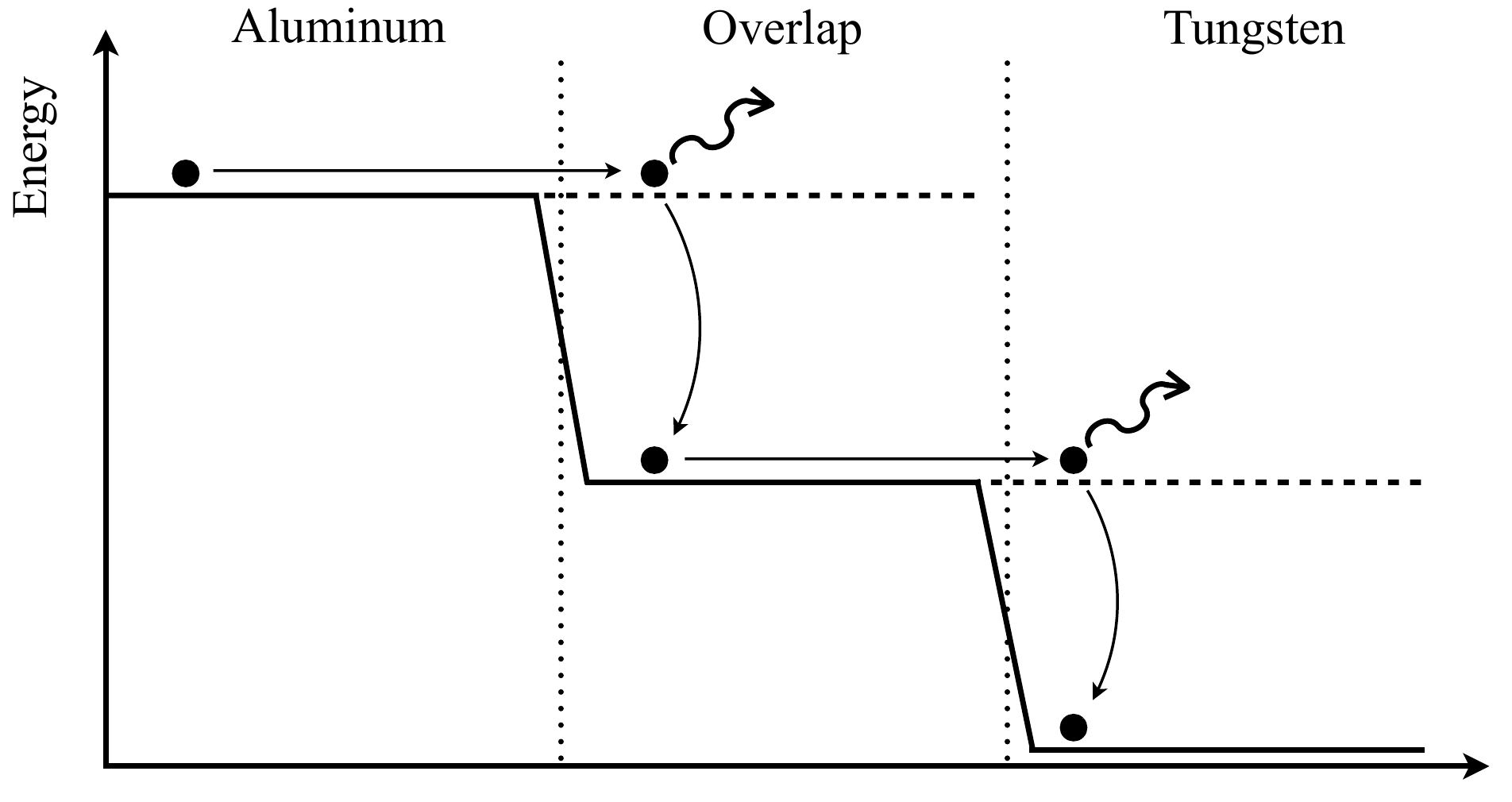}
\caption{Once quasiparticles are produced in the Al, they diffuse into the overlap bilayer region where the gap energy is smaller.  The quasiparticles then decay by releasing phonons, and they become trapped in the overlap region.  The same process occurs when the quasiparticles diffuse into W.\label{fig:GapDiagram}}.
\end{center}
\end{figure}

The SuperCDMS experiment searches for evidence of weakly-interacting massive particles (WIMPs) by detecting nuclear recoils in high-purity germanium detectors \cite{Akerib:2006}.  Nuclear recoils produce phonons and break electron-hole pairs in the Ge.  The phonons are detected at the top and bottom faces of the Ge crystals using tungsten-aluminum quasiparticle-trap-assisted electrothermal feedback transition-edge sensors (QETs) \cite{QET:1996}.  The phonons from the Ge substrate are collected by superconducting Al ``fins" that cover a substantial fraction of the crystal surface.  The absorbed phonons break Cooper pairs and produce quasiparticles.  Quasiparticles diffuse through the Al to the interface with the W TES, where W and Al overlap to form a bilayer, as shown in Figure \ref{fig:GapDiagram}.  The quasiparticle energy in the W is lower than the energy in the bilayer, which is in turn lower than in the Al.  As a result, the quasiparticles are ultimately trapped and release their energy in the bilayer region and the W TES.  The energy from the quasiparticles is then detected as a change in voltage across the TES due to an increase in temperature \cite{Irwin:1995}.  A large number of TESs are read out in parallel in each of the eight channels of a detector.

An important point is that some of the phonons in the Ge substrate, so-called athermal phonons, carry the position information of the initial event.  Thus, maximizing the detected signal from this phonon population is an important design goal.  Doing so requires maximizing the phonon collection area on the surface and maintaining sufficient bandwidth to resolve the signal.  Increasing the phonon collection area is achieved simply by increasing the number of TESs and their attached Al fins.  The bandwidth, on the other hand, is set by the $L/R$ time-constant that determines the frequency at which the detector response rolls off.  Since the inductance $L$ is fixed by the SQUIDs that are used to read out the TESs, the resistance of the TESs is set in the design to match the desired bandwidth.  In SuperCDMS detectors, the actual phonon channels each consist of many hundred TESs arranged in parallel.  In order to maintain the desired channel resistance as more TESs are added, the length of each TES must be increased.  Sufficiently long TESs can separate into two phases, degrading the signal to noise.  Thus, optimizing the detectors involves a trade-off between maximizing phonon collection and minimizing the phase separation that can result from using long TESs.

In this paper, we discuss a simple simulation of the SuperCDMS TESs which allows us to study phase separation in the detectors.  The model has several parameters: some that can be measured directly and others that must be tuned using diagnostic data.  We describe the effects of each model parameter on the simulation, and describe a procedure for tuning the simulation to better understand and predict detector performance.

\section{Simulation}
To simulate a TES that could be phase-separated, we model a single TES as a one-dimensional array of nodes connected in series \cite{Leman:2006}.  The length of a TES in a SuperCDMS detector is the distance between the first and last nodes of this one-dimensional array.  We let each node have its own temperature and resistance, so that the two ends of the TES can have different phases.

In the model and experiments, each TES can be characterized by several static and tunable parameters, as discussed below \cite{Leman:2011}.
\begin{figure}
\begin{center}
\includegraphics[scale=1]{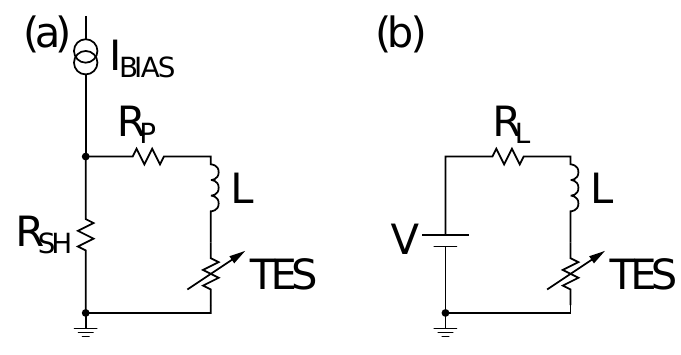}
\caption{Diagram of the TES circuit (a) and the Thevenin equivalent diagram (b).  From \cite{Irwin:2005}.\label{fig:TEScircuit}}
\end{center}
\end{figure}

\begin{enumerate}
\item \emph{Parasitic resistance} ($R_p$) - In the superconducting phase, the circuit has a residual parasitic resistance due to impurities and resistance in the leads that read out the TESs.
\item \emph{Normal resistivity} ($\rho_n$) - Each W TES has a characteristic normal resistance when it is just above the critical temperature, although small fabrication differences may cause this to be different for different TESs.  We denote the normal resistance of a TES by $R_n$.
\item \emph{Electron-phonon coupling parameter} ($\Sigma$) - This parameterizes the thermal coupling between the electron system and the phonon system within the W TES and W spurs that connect to the Al fins.  It has been measured to be approximately $3.2 \times 10^8$ Wm$^{-3}$K$^{-5}$, in the absence of the Al fins that connect to the Ge substrate and the TES \cite{Hart:2009}.  The geometry of these connections can vary somewhat between TESs, and the effect of this variation and the volume of Al material on $\Sigma$ of the TES is difficult to quantify \emph{a priori}.  We therefore allow this parameter to be tuned in order to match the simulation to diagnostic data.  
\item \emph{Specific heat capacity} ($C$) - The specific heat capacity of a TES used in SuperCDMS detectors ideally should be the value for W.  However, the presence of the Al fins changes the effective specific heat of the device in ways that are difficult to quantify \emph{a priori}.  As a result, we also allow this parameter to vary.  
\item \emph{Diffusion constant} ($D$) - This parameterizes the one-dimensional diffusion along the length of a TES.  Because of possible sub-micron-scale variations in the photolithographically patterned W TESs on the detectors, we treat this as a tunable parameter.  
\item \emph{Critical temperature} ($T_c$) - The critical temperature of a TES is the temperature of the superconducting transition, which may vary slightly from one TES to another.  This can be measured to within about 1~mK.  The critical current of a TES (the current needed to drive the TES normal) is measured at several different temperatures and the critical current vs. temperature data is extrapolated to zero current.  The temperature corresponding to zero current is taken to be the critical temperature of the TES.
\item \emph{Transition width} ($\Delta T_c$) - This describes the width of the superconducting transition, and it is related to the 10-90\% transition width by $\Delta T_{10-90} = 2 \Delta T_c \tanh^{-1} (0.8)$.  For the simulation, we assume that $\Delta T_{10-90} = 0.35$~mK, consistent with measurements.  This value is determined from resistance measurements.  
\item \emph{Spatial distribution of critical temperatures} (characterized by variance $\sigma_T^2$) - In the model, the $T_c$ of each TES is allowed to vary across the surface of the detector.  Due to the process by which the W is patterned on the Ge, small variations in the value of $T_c$ across the detector can occur.  In particular, small but nonzero radial gradients and linear gradients across the detector surfaces have been measured in some detectors.  Measurements of the detectors considered here indicated a small random variation of $T_c$ between individual TESs, but no systematic gradient.  For the purposes of this discussion, we therefore define the $T_c$ of each TES using a Gaussian with mean at the measured $T_c$ and a tunable variance $\sigma_T^2$.
\end{enumerate}

In our model, heat can be transferred to and from an individual TES by two processes: heat transfer to the substrate and Joule heating.  Radiative effects are negligible.  In order to model the heat transfer to the substrate, the functional form of the resistance of the TES as a function of temperature and current must be assumed.  We take $R_i(T_i,I_i)$ to have the following form, motivated by Ginzburg-Landau theory near $T_c$ \cite{Cabrera:2008},
\begin{equation}\label{eqn:Rfunc}
R_i(T_i,I_i) = \frac{R_n}{2} \left[ 1 + \tanh \left( \frac{T_i(t) - T_c}{\Delta T_c} + (I(t)/I_0)^{2/3} \right) \right],
\end{equation}
where $I_0$ is a constant, and the subscript $i$ labels the nodes into which each TES is subdivided.  $T_i(t)$ is the temperature of the electron system of the $i$th node at time $t$, and $I_i(t)$ is the current through the $i$th node at time $t$.  Joule heating and cooling of the TES to the substrate then gives the net power transfer to the TES,
\begin{eqnarray*}
P_i(t+1) &= &CV \frac{T_i(t+1) - T_i(t)}{\Delta t}\\
& = &V \Sigma (T_{bath}^5 - T_i(t)^5) + I^2 R_i(t),
\end{eqnarray*}
where $C$ and $V$ are the specific heat capacity and specific volume of W for a single TES node, $T_{bath}$ is the temperature of the constant temperature of the phonon system, $\Delta t$ is the time step of the simulation, and $\Sigma$ is the electron-phonon coupling parameter described above.  The first term gives the cooling to the Ge substrate and is motivated by \cite{Hart:2009}.  The second term gives the Joule heating.

Additionally, heat can be transferred between neighboring nodes within one TES by diffusion.  This process is described by a discrete 1D diffusion equation
\begin{equation}\label{eqn:Diffusion}
D \frac{T_{i+1}(t) - 2T_i(t) + T_{i-1}(t)}{(\Delta x)^2} = \frac{T_i(t) - T_i(t-1)}{\Delta t},
\end{equation}
where $D$ is the diffusion constant, and $\Delta x$ is the width of each node.

Each SuperCDMS detector has eight independent phonon channels (four on the top surface and four on the bottom surface), each of which contains $\sim500$ TESs in parallel and wired in a constant-voltage configuration. Real data obtained with a SuperCDMS detector therefore reflects the ensemble response of many different TESs. However, given the uniformity of the patterned TESs and the quality of the data observed, the parallel array has been shown to behave qualitatively the same as a single TES.

\section{Tuning with $I_b$-$I_s$ curves}
To use the simulation that we developed, the parameters described in the previous section must be tuned to data.  A bias current-sensor current ($I_b$-$I_s$) curve is one useful test that we perform in the lab, which allows us to tune $R_p$, $R_n$, $\Sigma$, and $\sigma_T^2$.  This test involves applying a linearly decreasing bias current to the TES circuit and measuring the current response.  Note that the TES itself is voltage biased, but we actually monitor the bias current through the parallel TES circuit (see Figure \ref{fig:TEScircuit}).  The $I_b$-$I_s$ curve is therefore equivalent to an $I$-$V$ curve.  The rate of decrease in current is chosen to be slow compared to all time scales of the detector response so that the TES is approximately in thermal equilibrium with the Ge substrate at all times during the test.  Since the device is in thermal equilibrium, we can recast the $I_b$-$I_s$ data into a plot of power dissipation as a function of applied bias current, shown in Figure \ref{fig:SampleIbPower}.

\begin{figure}
\begin{center}
\includegraphics[scale=0.5]{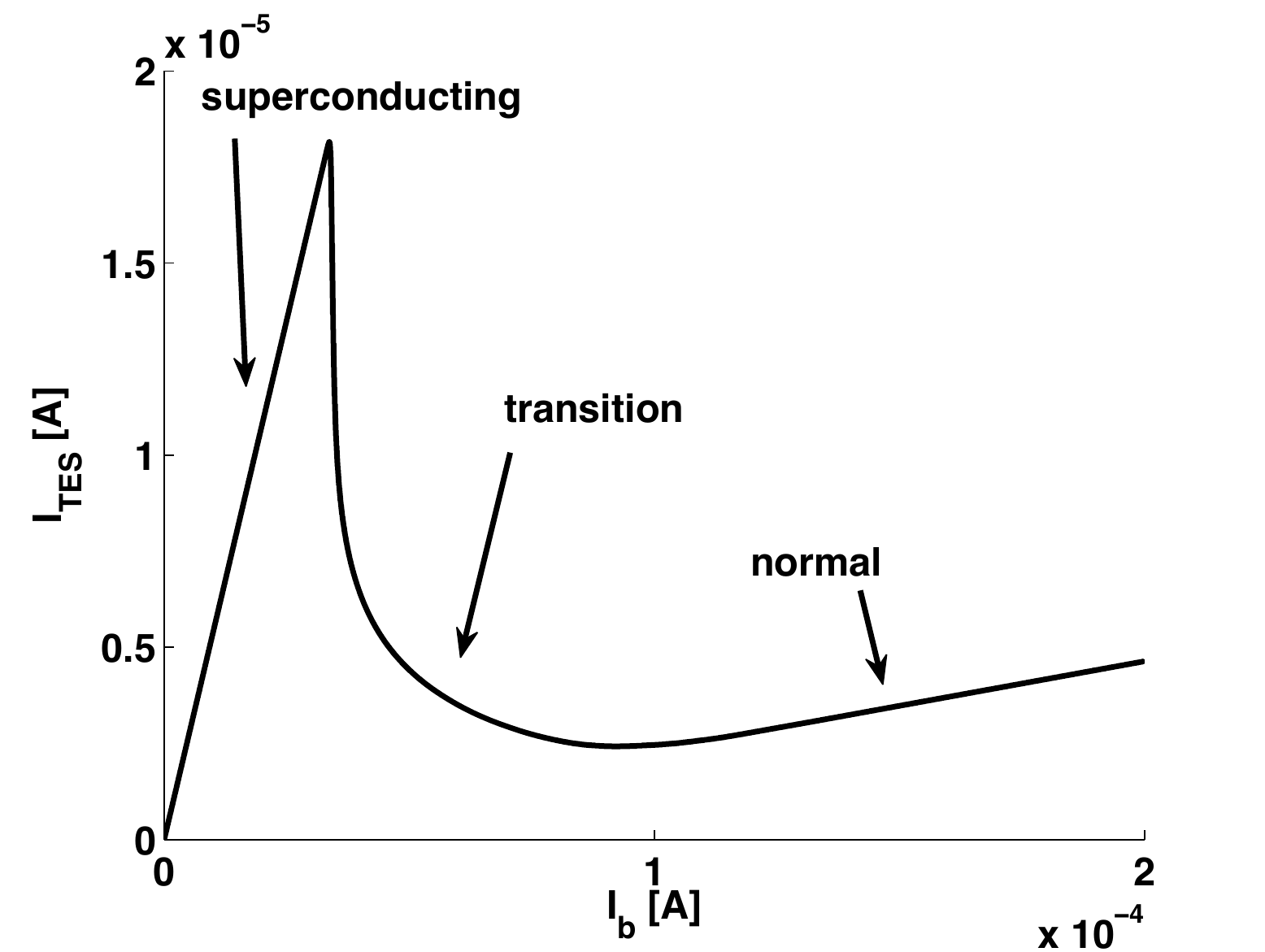}
\caption{Current response in the TES due to a slowly decreasing bias voltage in the phase-uniform case.\label{fig:SampleIbIs}}
\end{center}
\end{figure}

\begin{figure}
\begin{center}
\includegraphics[scale=0.5]{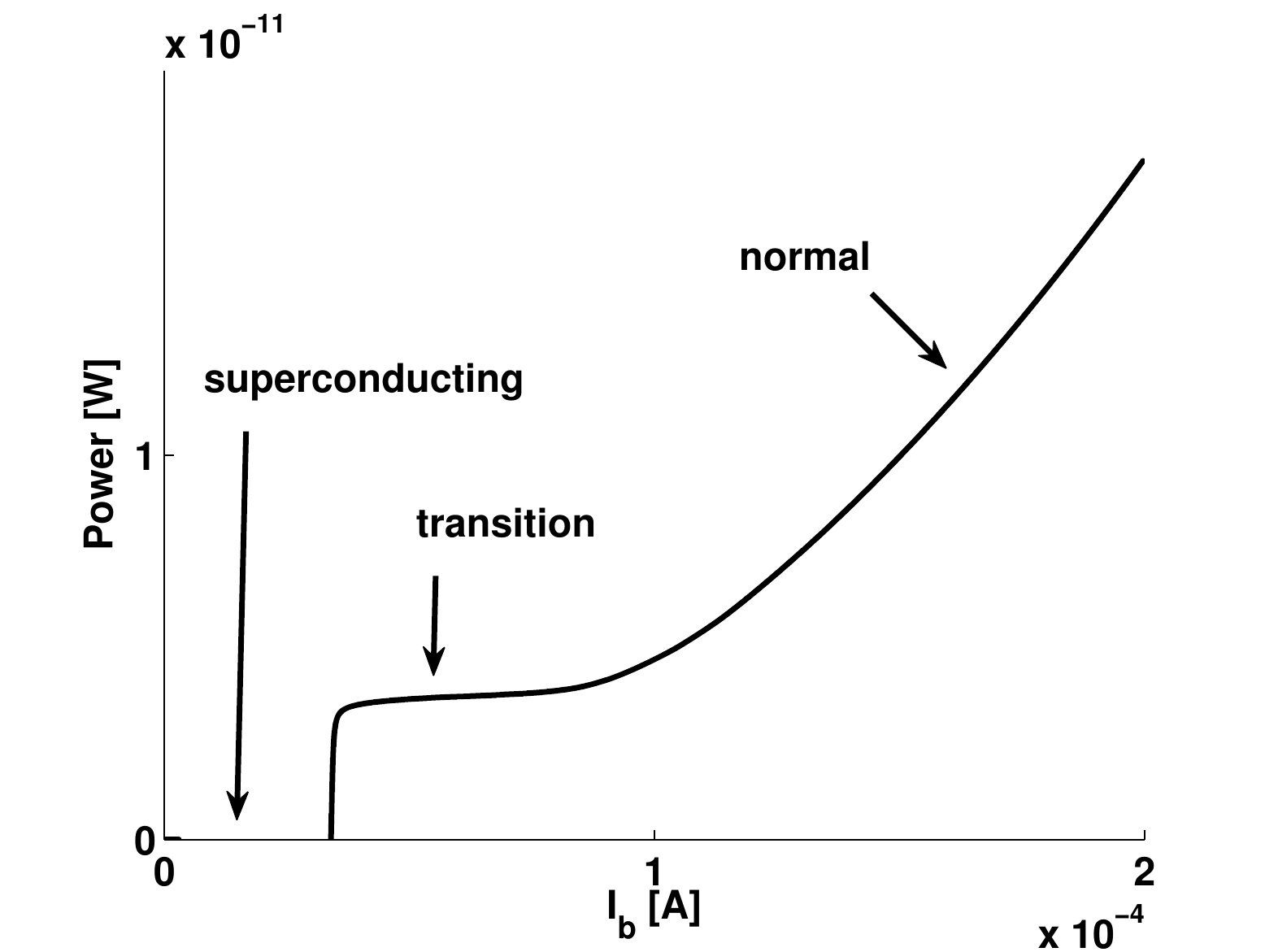}
\caption{Power dissipated through the TES as a function of a slowly decreasing bias voltage in the phase-uniform case.\label{fig:SampleIbPower}}
\end{center}
\end{figure}

At high applied bias current (e.g. 0.3~mA), a channel of TESs has an approximately constant resistance from the parasitic resistance ($\sim$15m$\Omega$) and the normal resistance of the W ($\sim$0.7$\Omega$), and it obeys Ohm's law.  Similarly, at low current (e.g. 0.02~mA), the TES obeys Ohm's law and has an approximately constant resistance given by the parasitic resistance alone.  This is evident from the linear regions of Figure \ref{fig:SampleIbIs}.  As the bias current decreases from high current and the TES begins to enter the transition, the TES temperature is stabilized by electrothermal feedback and slowly progresses through the transition.  As the temperature decreases, the resistance decreases, and the current through the TES increases, producing the spike in Figure \ref{fig:SampleIbIs}.  Since the TES is in equilibrium and its temperature is nearly constant in the transition, the power dissipated to the substrate is constant, producing the plateau seen in Figure \ref{fig:SampleIbPower}.

\begin{figure}
\begin{center}
\includegraphics[scale=0.5]{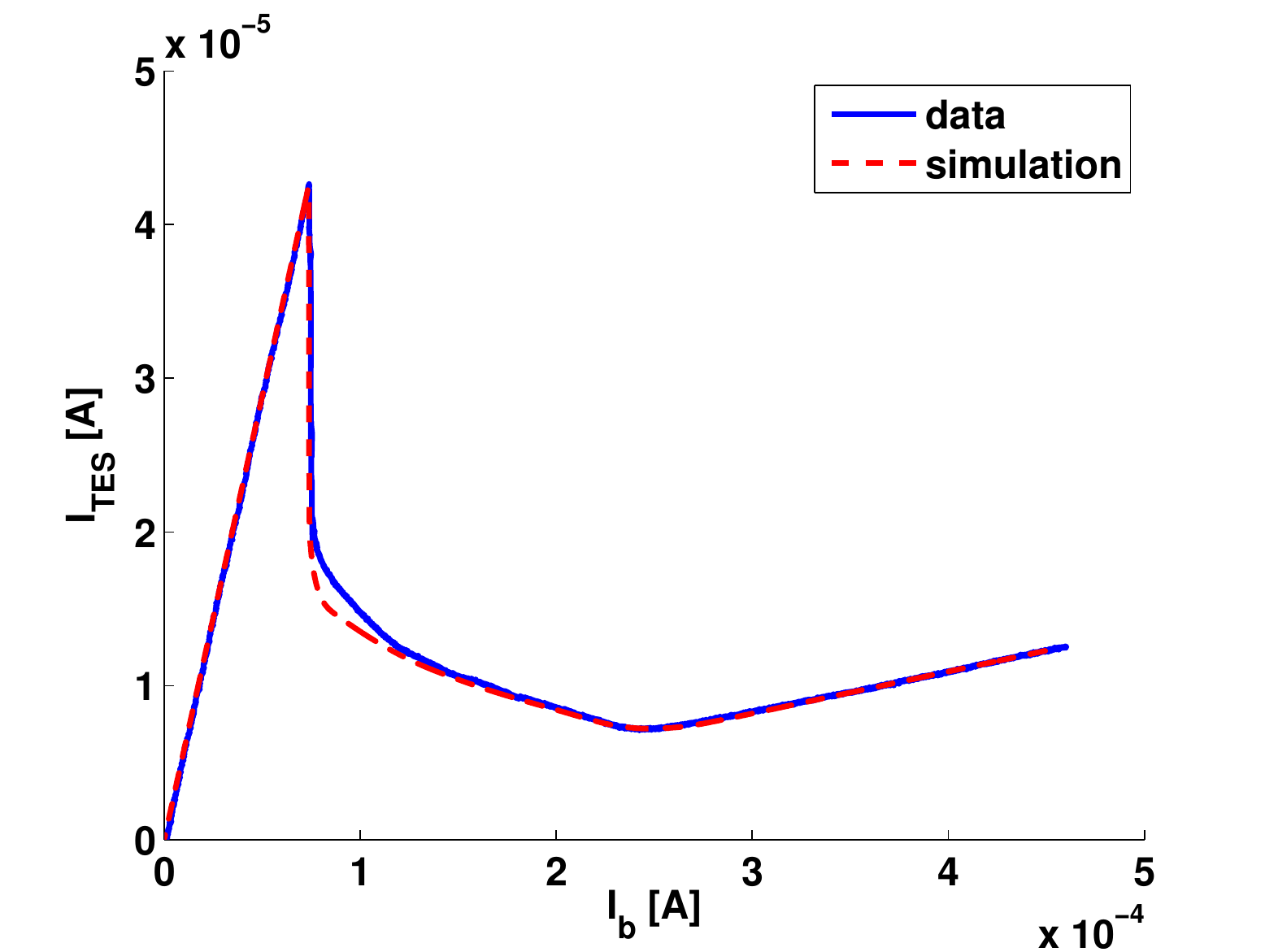}
\caption{Experimental results (solid blue) and simulation (dashed red) for the current response of a phase-separated TES as a function of the TES bias current.\label{fig:SampleIbIs_PhaseSep}}
\end{center}
\end{figure}

\begin{figure}
\begin{center}
\includegraphics[scale=0.5]{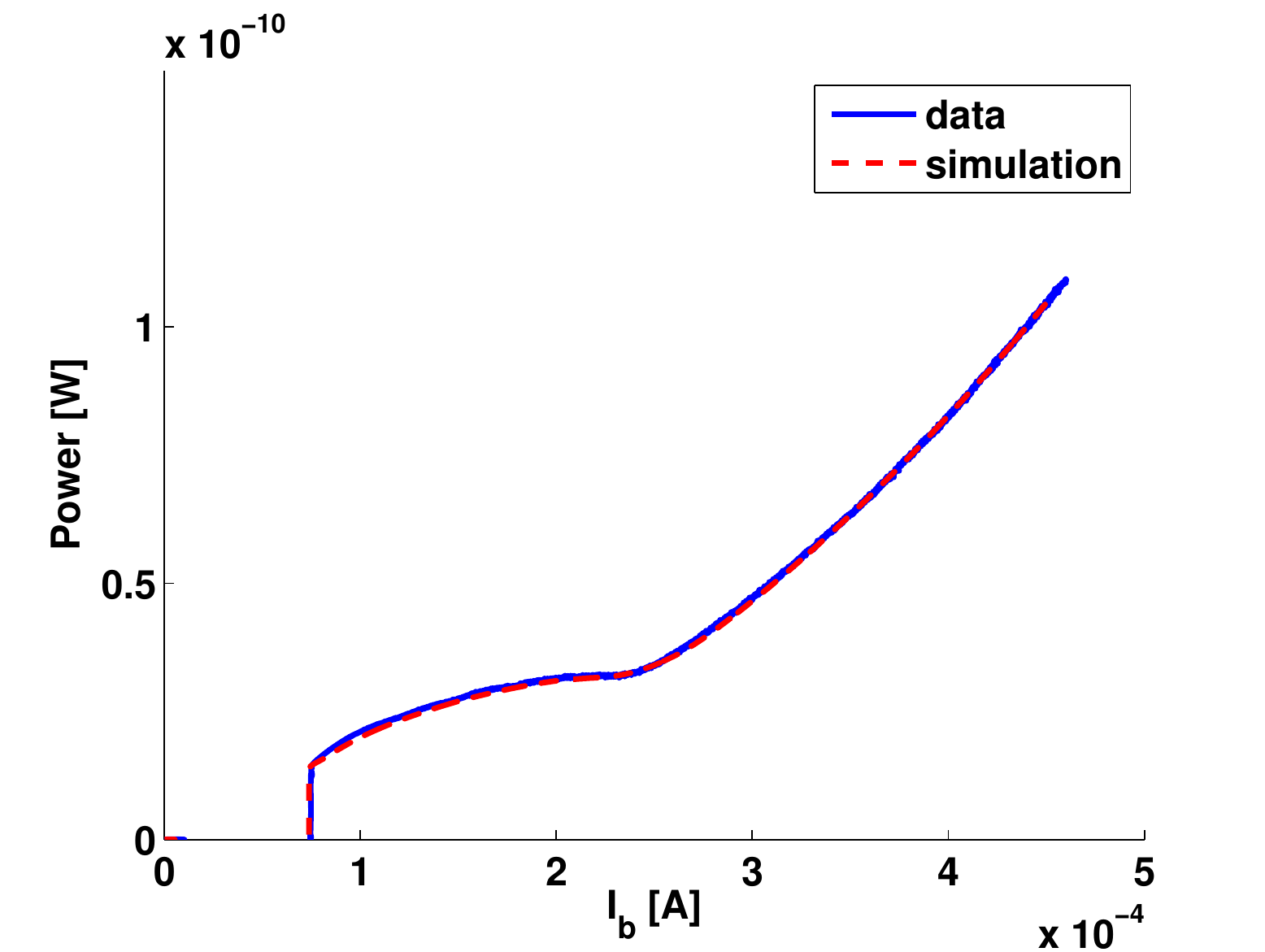}
\caption{Experimental results (solid blue) and simulation (dashed red) for the power dissipated by a phase-separated TES as a function of TES bias current.\label{fig:SampleIbPower_PhaseSep}}
\end{center}
\end{figure}

In the phase-separated case, part of the TES is the normal phase and another part is the superconducting phase.  When this occurs, the shape of the $I_b$-$I_s$ curve changes in the transition, as shown in Figure \ref{fig:SampleIbIs_PhaseSep}.  The plateau region of the power curve becomes considerably more sloped, as shown in Figure \ref{fig:SampleIbPower_PhaseSep}.  At some point in the transition the TES switches from a phase-uniform to phase-separated state.  In the phase-separated state, a significant fraction of the TES is at fixed resistance--either the normal resistance or nearly zero resistance.  As the bias current decreases, the power dissipation also decreases.

To tune $R_p$, $R_n$, $\Sigma$, and $\sigma_T^2$ with these data, we attempt to find the parameters that best match the bias current versus power curve for simulation and data (e.g. Figure 4).  First, the parasitic and normal resistance are determined by performing a linear fit to the normal and superconducting regions of the $I_b$-$I_s$ curve and extracting the slope of each, as shown in Figures \ref{fig:SampleIbIs} and \ref{fig:SampleIbIs_PhaseSep} for example.  The parameters $\Sigma$ and $\sigma_T^2$ are tuned by running the simulation for a matrix of parameter values.  The best-fit values of the parameters are selected by the least-squares difference between the data and simulated $I_b$-$I_s$ curves in the transition region.  This process is iterated to obtain the optimal parameters.

Graphically, increasing $\Sigma$ or $T_c$ has the effect of raising the flat region of the plot of bias current versus power dissipated.  This occurs because the heat transfer to the substrate is larger and the transition is reached at a higher bias current.

The data from $I_b$-$I_s$ curves alone cannot break all of the degeneracies between the model parameters.  There is considerable degeneracy between $\Sigma$ and $T_c$, for example.  Additionally, $C$ and $D$ have only a small effect on the shape of the $I_b$-$I_s$ curve, and this method is consequently fairly insensitive to these parameters.  For this reason, we take the approximate $T_c$ from critical current measurements, and we do not tune $C$ or $D$ with the $I_b$-$I_s$ curves.  

While the model fit to $I_b$-$I_s$ curves is generally good, it should be noted that systematic error is introduced through our assumption that the form of the resistance curve in the transition is given by equation (\ref{eqn:Rfunc}).  This is a phenomenological parameterization that produces acceptable results, but it does not represent the true form of the resistance curve.  Differences between this form and the true resistance curve contribute to the remaining discrepancy between simulated and measured curves.

\section{Tuning with sudden impulse}
\begin{figure}
\begin{center}
\includegraphics[scale=0.5]{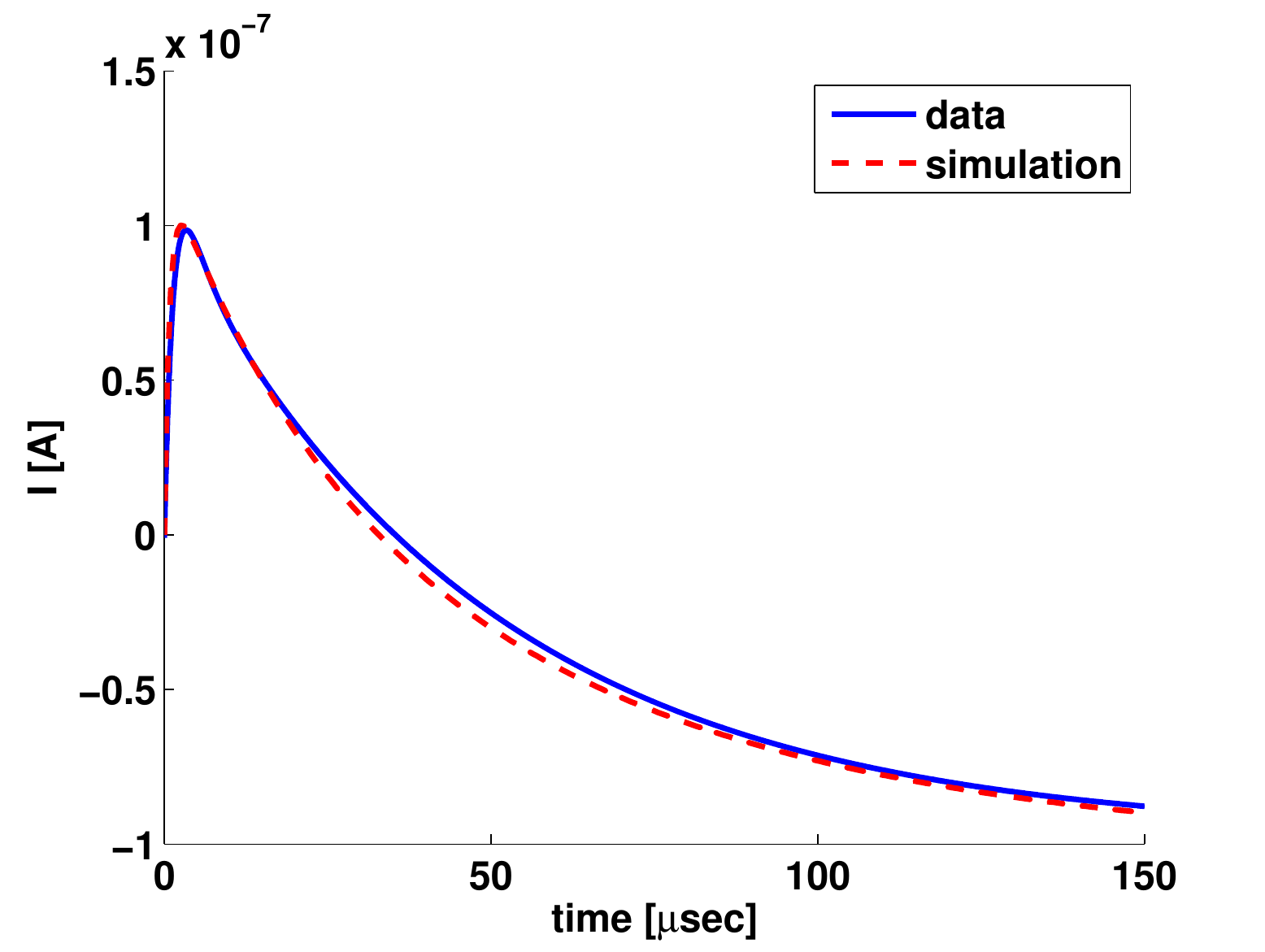}
\caption{Experimental (solid blue) and simulated (dashed red) TES response to a square wave bias current.\label{fig:SampleSquareWave}}
\end{center}
\end{figure}

To better tune the diffusion constant and the specific heat capacity, we study the detector response to a square wave current.  We apply a square wave of 0.025 mA peak-to-peak amplitude on top of a constant bias current of 1.54 mA.  A characteristic response associated with the rising edge of a square wave is shown in Figure \ref{fig:SampleSquareWave}.  The sudden increase in the voltage across the TES initially increases the current with a characteristic time scale $L/R$.  At slower time scales, the TES temperature and its resistance increase through Joule heating, which decreases the current flow.  Eventually a new quasi-equilibrium point is reached with both smaller current and larger resistance than originally.

The specific heat controls the temperature change needed to restore internal equilibrium and the rate at which energy is exchanged to the substrate, while the diffusion constant affects the time to reach internal thermal equilibrium along the TES.  For these reasons, the specific heat and diffusion constants strongly affect the decay time of the TES response to a square wave, and both parameters can be tuned by fitting data and simulation.  The parameter $\Sigma$ also has a significant effect on the decay time of the current response by controlling the rate at which the TES reaches thermal equilibrium with the substrate.

We find, in particular, that increasing the diffusion constant tends to decrease the decay time.  Increasing the specific heat, on the other hand, increases the decay time.  A larger specific heat implies that the temperature excursions due to the changing bias current are smaller; therefore, the cooling power will be smaller and the time to return to equilibrium will be larger.  The difference between the sensor current before the square wave edge and at a long time after is determined by the overall constant bias current applied: being biased at different points in the superconducting transition will result in different changes in the resistance of the TES when a square wave bias is applied.

\begin{table}
\begin{center}
\begin{tabular}{l | c}
Parameter & characteristic value \\ \hline
mean $T_c$ & 90 mK  \\
$\Delta T_c$ & 0.35 mK  \\
$\sigma_T$ & 1 mK \\
$\Sigma$ & $4.8 \times 10^8$ W $\Omega$ K$^{-5}$  \\
$\rho_n$ & $1.2 \times 10^{-7}$ $\Omega$ m \\
$D$ & $4.03 \times 10^{-4}$ m$^{-2}$ s \\
$c_p$ & $37.0$ J K$^{-1}$ m$^{-3}$
\end{tabular}
\caption{Characteristic model parameters found from matching of $I_b$-$I_s$ curves and square-wave response.\label{tab:characteristicParams}}
\end{center}
\end{table}

The response of a phase-uniform TES is characterized by a single exponential risetime and a single exponential falltime \cite{Irwin:2005}.  In contrast, the response of a phase-separated TESs is best described by a single exponential risetime and two exponential falltimes.  The qualitative behavior of these pulses is well-reproduced by the simulation, as seen in Figure \ref{fig:SampleSquareWave}, although it is difficult to precisely match the parameters of the exponentials describing the data and simulated pulses.

\section{Phase-Separation Length}
The phase-separation length for a TES is the maximum length of a TES before it becomes phase-separated.  There is a simple analytic expression for this length, assuming the Wiedemann-Franz law \cite{Hart:2009},
\begin{equation}\label{eqn:analyticPhaseSepL}
\ell = \sqrt{\frac{\pi^2 L}{\alpha \Sigma T_c^3 \rho_n}},
\end{equation}
where $L = 2.44 \times 10^{-8}$ W $\Omega$ K$^{-2}$ is the Lorentz constant, $\rho_n$ is the normal resistivity, and $\alpha = \partial \log R / \partial \log T$.  This expression is not exact because Wiedemann-Franz holds only approximately as these temperatures.  For the functional form of the resistance function that we assume in equation (\ref{eqn:Rfunc}), we have
\begin{equation}
\alpha(u) = \frac{T}{\Delta T_c} \frac{1}{ (\cosh^2 u)(1+\tanh u)},
\end{equation}
where
\begin{equation}
u = \frac{T - T_c}{\Delta T_c} + (I/I_0)^{2/3},
\end{equation}
and $T$ and $I$ are the temperature and current at the bias point.  For some characteristic parameter values shown in Table \ref{tab:characteristicParams}, we find that equation (\ref{eqn:analyticPhaseSepL}) gives a phase-separation length $\ell = 120$ $\mu$m.

\begin{figure}
\begin{center}
\includegraphics[scale=0.5]{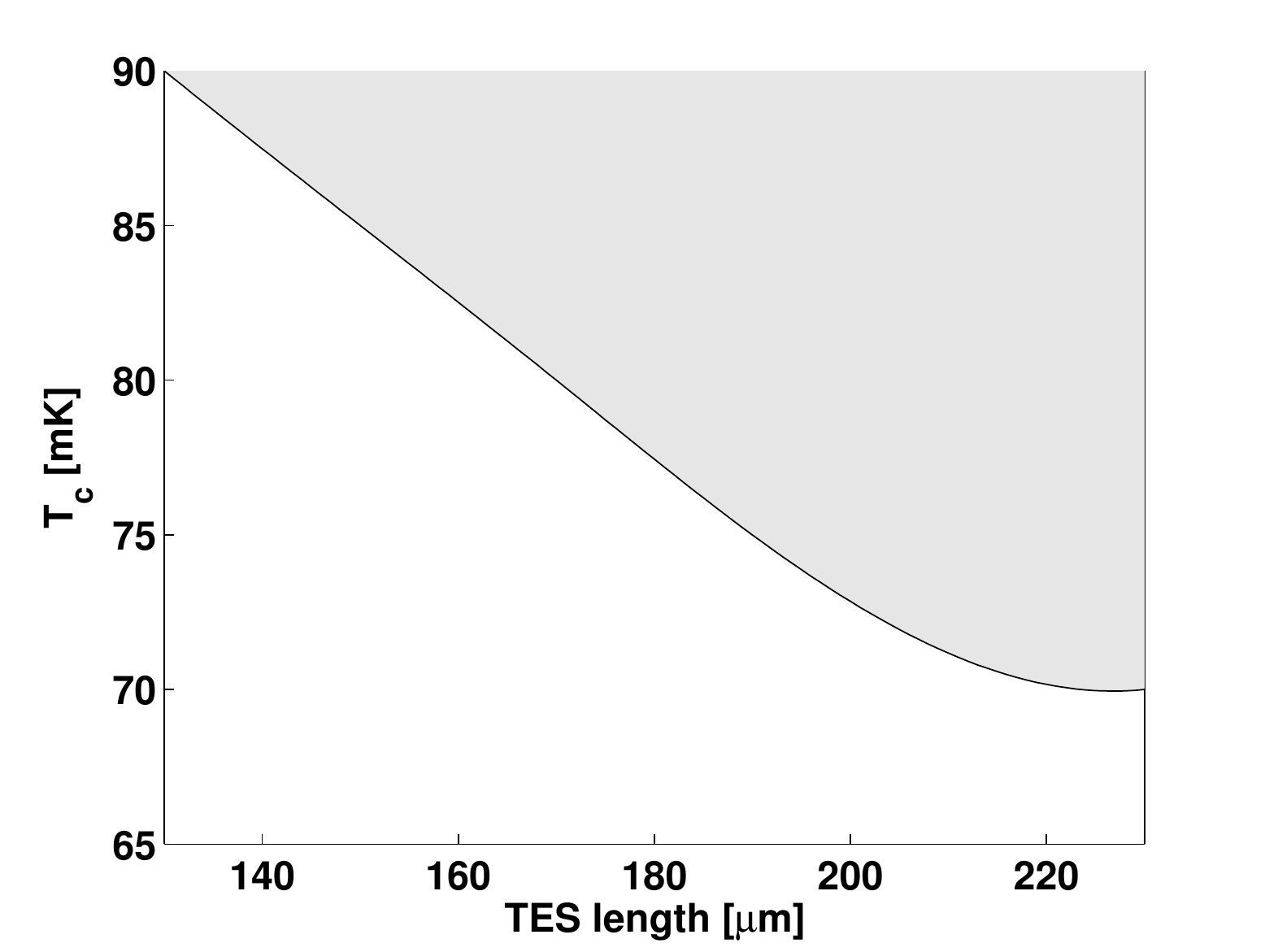}
\caption{Simulated phase-separation boundary.  The shaded points above the line are combinations of TES length and $T_c$ that produce phase-separation when biased at 1/3 of the normal resistance.  Points below the curve produce phase-uniform TES. \label{fig:PhaseSepLengthSurfacePlot}}
\end{center}
\end{figure}

We can also use the simulation to obtain an estimate of the phase-separation length.  To do this, we run a simulation of 100 TESs in parallel at a constant bias current, and count the number of channels with phase-separated TESs.  Finally, since the phase separation depends on both $T_c$ and the length of the TES, we can use the simulation to map out regions where the TESs are phase-separated or phase-uniform.  This is shown in Figure \ref{fig:PhaseSepLengthSurfacePlot}.  The result is in general agreement with equation (\ref{eqn:analyticPhaseSepL}).  It is important to note, however, that Figure \ref{fig:PhaseSepLengthSurfacePlot} assumes a fixed value of $\Delta T_c$ for all values of $T_c$ and TES lengths, so each point on the curve corresponds to a different value of $\alpha$.

\section{Effect on Phonon Simulation}
\begin{figure}
\begin{center}
\subfigure[Comparison of simulated (dashed red) and data (solid blue) phonon pulses in a SuperCDMS detector, using the tuned TES parameters.  Different panes correspond to different channels of the detector. ]{
   \includegraphics[scale =0.6] {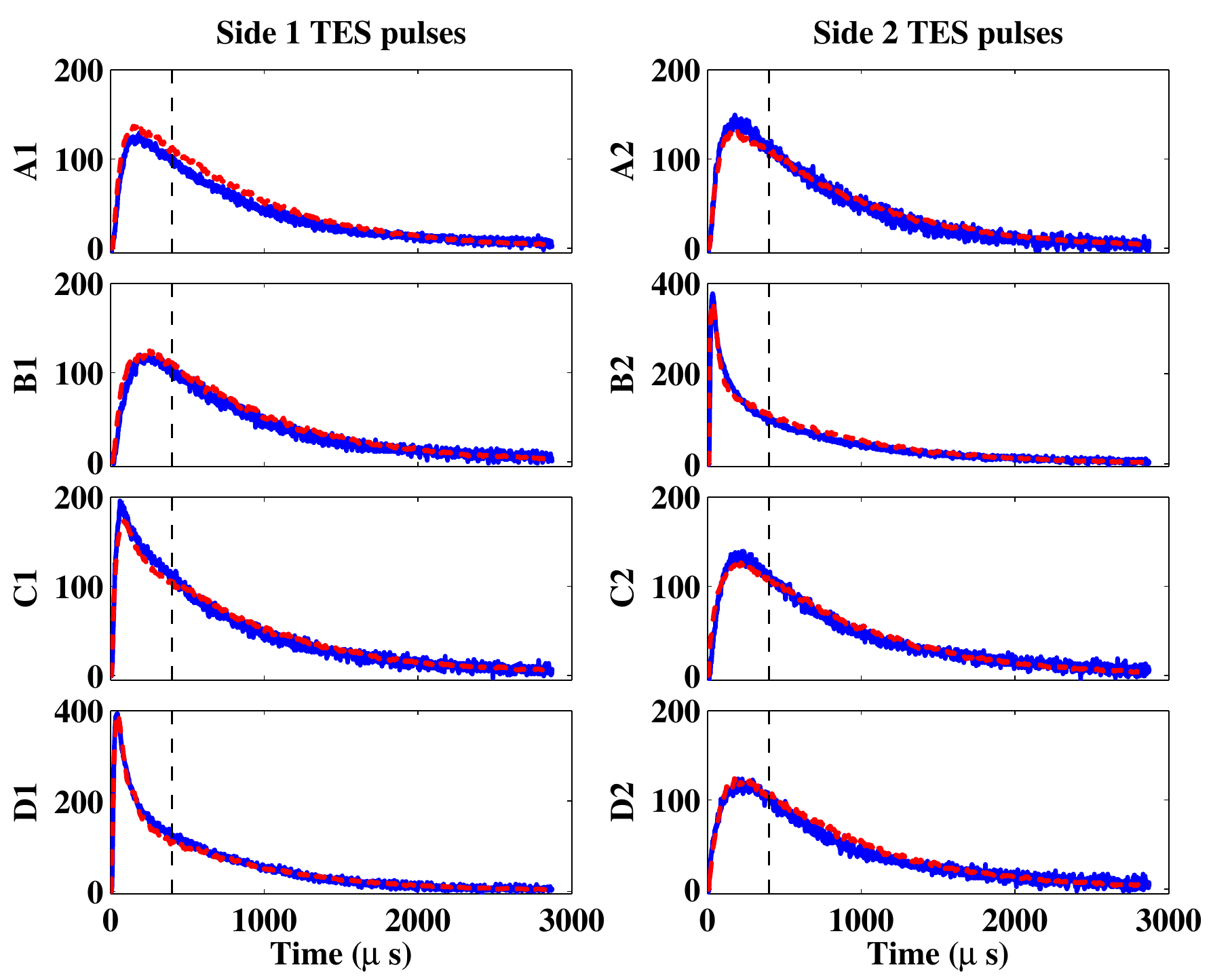}
   \label{fig:PhononPulses1}
 }
 
 \subfigure[The first 400~$\mu$s of each phonon pulse in (a), showing good agreement in the rising edge.]{
   \includegraphics[scale =0.6] {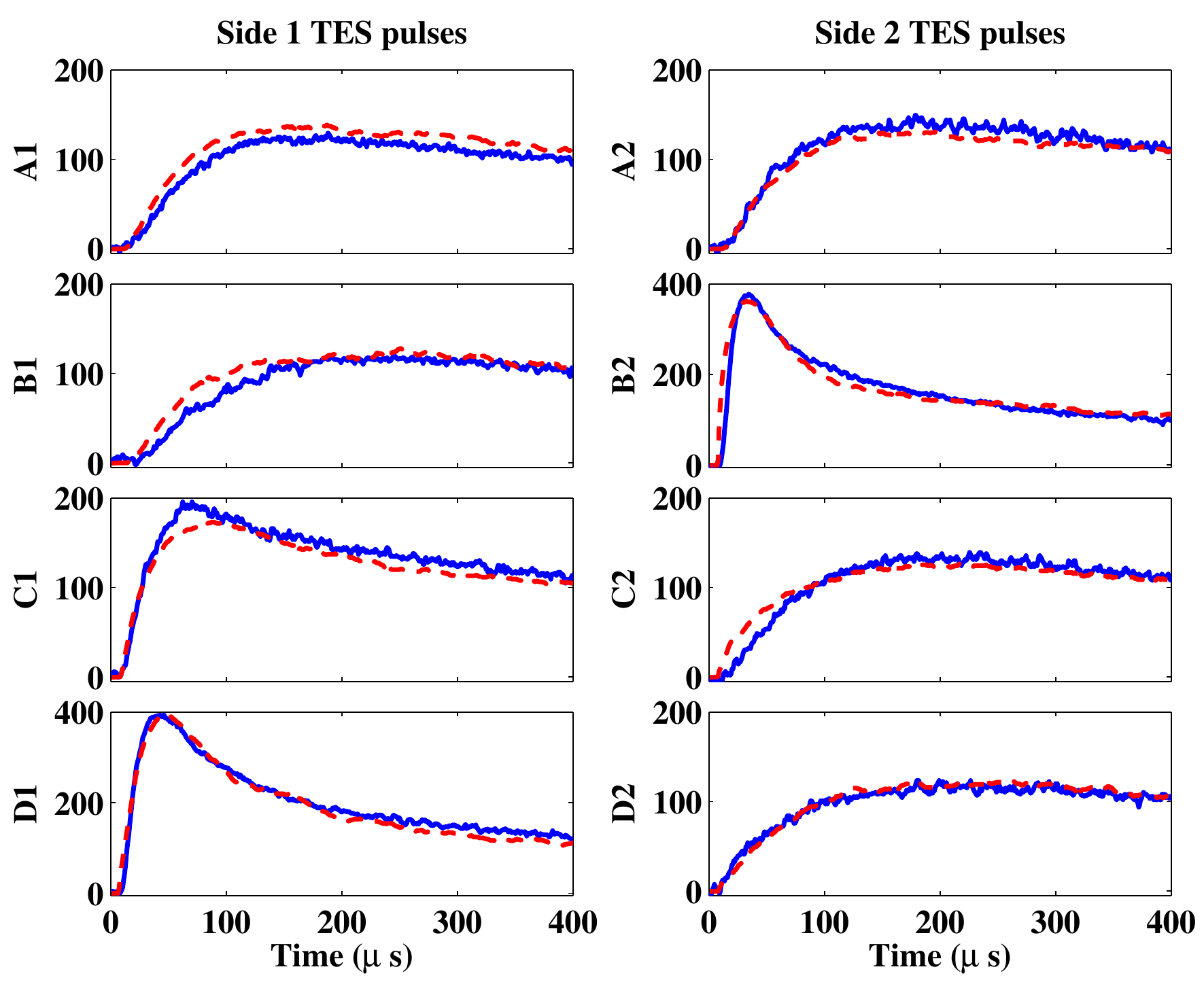}
   \label{fig:PhononPulses2}
 }
\end{center}
\caption{Phonon pulses in data and simulation.}
\end{figure}

The TES simulation is part of a broader Monte Carlo simulation of the SuperCDMS detectors, which models the full charge and phonon physics of each event \cite{DMCGeneral,Leman:2011}.  Phonons are parsed into individual TES, from which pulses are constructed.  Modeling the TES using correct parameters is important for producing realistic phonon pulses.  The effect of this tuning procedure is depicted in Figures \ref{fig:PhononPulses1} and \ref{fig:PhononPulses2}, which show some phonon pulses from data and the best simulated matches from simulated template pulses.  Before the tuning of TES parameters, there was significant disagreement between data and simulation at short timescales.  After the correct parameters are used, the agreement is considerably improved.

\section{Conclusion}
We have described a simple numerical simulation of phase separation in TESs as used in the SuperCDMS experiment.  The simulation reproduces the general behavior of $I_b$-$I_s$ curves for $\sim500$-element TES arrays.  The simulation also correctly reproduces the current response to square wave bias voltage, and it can be used to estimate the phase separation length at various critical temperatures and TES lengths.  This agreement is useful for tuning the simulation parameters.

This simulation has been integrated into the broader Monte Carlo simulation of SuperCDMS detectors.  Proper TES modeling has been shown to be important for correctly reproducing the behavior of phonon pulses on short timescales.  More broadly, the TES simulation is an important tool for optimizing the design future TESs and detectors.

\section{Acknowledgements}
The authors wish to acknowledge support by the United States National Science Foundation under Grant No. PHY-0847342 and PHY-0801712.


\begin{thebibliography}{9}
\bibitem{CDMSScience:2010}
Z.~Ahmed, et al., Science, {\bf 327}, 1619 (2010).

\bibitem{CRESST:2009}
G.~Angloher et al., Astropart. Phys., {\bf 31}, 270 (2009). 

\bibitem{microx:2009}
M.~E. Eckart et al., AIP Conf. Proc., {\bf 1185}, 699 (2009).

\bibitem{IXO:2010}
J.~W.~den Herder et al., Proc. SPIE, {\bf 7732}, 77321H (2010).

\bibitem{SPT:2004}
J.~E.~Ruhl et al., Proc. SPIE, {\bf 5498}, 11 (2004).

\bibitem{SPIDER:2008}
C.~L.~Kuo, et al., Proc. SPIE, {\bf 7020}, 70201I (2008).

\bibitem{IrwinAPL:1995}
K.~D.~Irwin, Appl. Phys. Lett., {\bf 66}, 1998 (1995).

\bibitem{Irwin:2005}
K.~D.~Irwin, G.~C.~Hilton, in: C. Enss (Ed.), {\it Cryogenic Particle Detection}, Springer-Verlag, Berlin, 2005, pp. 63-149.

\bibitem{Leman:2006}
S.~W.~Leman, PhD dissertation, Stanford University (2006).

\bibitem{AndersonLTD:2011}
A.~J.~Anderson et al., Proc. LTD-14, J. Low Temp. Phys., submitted (2011).

\bibitem{Galeazzi:2011}
M.~Galeazzi, IEEE Trans. Appl. Supercond., {\bf 21}, 267 (2011).

\bibitem{Akerib:2006}
D.~S.~Akerib, et al., Nucl. Instrum. Meth. A, {\bf 559}, 411 (2006).

\bibitem{QET:1996}
S.~W.~Nam, et al. Nucl. Instrum. Meth A. {\bf 370}, 187 (1996).

\bibitem{Irwin:1995}
K.~D.~Irwin, et al., Rev. Sci. Instrum., {\bf 66}, 5322 (1995).

\bibitem{Leman:2011}
S.~W.~Leman, 1109.1193, (2011).

\bibitem{Hart:2009}
S.~J.~Hart, et al., AIP Conf. Proc., {\bf 1185}, 215 (2009).

\bibitem{Cabrera:2008}
B.~Cabrera, J. Low Temp. Phys., {\bf 151}, 82 (2008).

\bibitem{DMCGeneral}
S.~W.~Leman, et al., in preparation.
\end{thebibliography}
\end{document}